# Quantification and correction of systematic errors due to detector time-averaging in single molecule tracking experiments

Nicolas Destainville,* and Laurence Salomé†

*Laboratoire de Physique Théorique, UMR CNRS-UPS 5152, Université Paul Sabatier, 31062 Toulouse, France ; † Institut de Pharmacologie et de Biologie Structurale, UMR CNRS-UPS 5089, 205, route de Narbonne, 31077 Toulouse, France

ABSTRACT   Single molecule tracking is a powerful way to look at the dynamic organization of plasma membranes. However there are some limitations to its use. For example, it was recently observed (1), using numerical simulation, that time-averaging effects inherent to the exposure time of detectors are likely to bias the apparent motion of molecules confined in microdomains. Here, we solve this apparently limiting issue analytically. We explore this phenomenon by calculating its effects on the observed diffusion coefficients and domain sizes. We demonstrate that the real parameters can be easily recovered from the measured apparent ones. Interestingly, we find that single molecule tracking can be used to explore events occurring at a time scale smaller than the exposure time.

It has been pointed out (1) that time-averaging, due to the exposure time of detectors in single molecule tracking experiments, can have dramatic effects. This is particularly important in measures of the apparent motion of tracked molecules (proteins or lipids) at the cell surface when they are confined in small regions of the membrane, such as rafts, synapses or other signaling platforms. Diffusion coefficients and size of confining domains can be significantly under-estimated. However, the arguments used relied on numerical simulation (1) and it seems important to validate them by analytical calculation. The work presented here addresses this issue and also enables the prediction of the range of experimental parameters within which this detector time-averaging effect perturbs the observations. Using systematic quantification of the time-averaging effects in the ranges of parameters of experimental relevance, we demonstrate that the values of diffusion coefficients or domain sizes are not significantly affected in a broad range of parameters. In addition, we show that these effects can be easily corrected and that real parameters can be recovered from those measured using simple formulae.

Consider a molecule diffusing in the membrane with diffusion coefficient D. Its displacements are followed by single molecule tracking by means of a detector with exposure time T. $\mathbf{R}_m(t)$ denotes the *measured* position of the molecule at time t (multiple of T). The *measured* mean-square deviation $MSD_m(t)$ is

$$MSD_m(t) = \langle (\mathbf{R}_m(s+t) - \mathbf{R}_m(s))^2 \rangle \quad (1)$$

where the brackets denote a discrete average over frames s. We further suppose that the diffusion is restricted to a square domain of side L. The usual experimental procedure to extract information from $MSD_m(t)$ is to fit it with the expected generic expression $MSD_m^{fit}(t)$ for a confined diffusion:

$$MSD_m^{fit}(t) = \frac{L_m^2}{3}\left(1 - \exp\left(-\frac{t}{\tau_m}\right)\right) \quad (2)$$

from which one extracts the *measured* (or apparent) domain size $L_m$ and equilibration time $\tau_m$. The *measured* diffusion coefficient $D_m$ is given by $D_m = L_m^2 / 12\tau_m$.

The *real* time-dependent positions of the molecule (as opposed to those *measured*) are denoted by $\mathbf{r}(t)$. Since the molecule is confined, they are correlated. The *real* equilibration time $\tau$ in the box is the typical decay-time of the following two-time correlator C(t) where averages are over times s:

$$C(t) \equiv \langle \mathbf{r}(s+t)\mathbf{r}(s) \rangle - \langle \mathbf{r}(s+t) \rangle \langle \mathbf{r}(s) \rangle \quad (3)$$
$$\cong Const. \exp(-t/\tau)$$

In practice, there are several time scales because the different modes of the diffusion operator do not decay at the same rate in the square box (2). The slowest mode decays exponentially with a decay-time $\tau_0 = L^2/(\pi^2 D)$ and the next modes have decay-times $\tau_0/(2k+1)^2$ with k an integer. Keeping the first-order term in the exact expansion of C(t) (2)

$$C(t) = \frac{16L^2}{\pi^4} \sum_{k=0}^{\infty} \frac{1}{(2k+1)^4} \exp\left(-(2k+1)^2 \frac{t}{\tau_0}\right) \quad (4)$$

leads to the approximation

$$C(t) \cong \frac{L^2}{6} \exp(-t/\tau) \quad (5)$$

as C(0), the variance of **r**(t), equals $L^2/6$. To restore the fact that C(t)–C(0) =1/2 $\langle(\mathbf{r}(s+t) - \mathbf{r}(s))^2\rangle$ = 2 D t at small t for a diffusing molecule, we need to set

$$\tau = L^2/12D. \quad (6)$$

Note that Eqs. 5 and 6 remain valid if the confining domain is not a square, but a disk, an ellipse or any more complex shape. In such cases, $\tau$ still represents the equilibration time and L is the typical domain size. For example, if the domain is a circle, its diameter is related to L by d= $(2/\sqrt{3})$ L. For a quadratic confining potential, $U(r)=1/2Kr^2$, Eq. 5 is even exact (5) and L is the typical width of the trap at temperature $\theta$ given by $L^2= 6k_B\theta/K$.

The *measured* position $\mathbf{R}_m(t)$ is related to the real one by

$$\mathbf{R}_m(t) = \frac{1}{T} \int_t^{t+T} \mathbf{r}(u) \, du. \quad (7)$$

After replacing $\mathbf{R}_m(t)$ using Eq. 7, the expansion of Eq. 1 leads to four correlators. Approximating them with Eq. 5 and setting x = $\tau$ / T gives

$$MSD_m(t) = \quad (8)$$
$$\frac{L^2}{3}\left[2x - 2x^2\left(1-e^{-\frac{1}{x}}\right) - \exp\left(-\frac{1}{x}\frac{t}{T}\right)x^2\left(e^{\frac{1}{x}} + e^{-\frac{1}{x}} - 2\right)\right]$$

This exact expression of the *measured* $MSD_m(t)$ in the case of a detector time-averaging is valid only if t ≥ T. If t < T, $MSD_m(t)$ is still calculable but this is beyond the scope of this Letter.

We extract the *measured* parameters from $MSD_m(t)$ as described above and compare them to the *real* ones. The domain size $L_m$ is obtained by equaling the large t limits of Eqs. 8 and 2:

$$L_m = L\left(2x - 2x^2\left(1-e^{-1/x}\right)\right)^{1/2} \equiv (g(x))^{1/2}. \quad (9)$$

From Eq. 2, we get $\tau_m$ *via* the simple relation

$$MSD_m(\tau_m) = L_m^2/3 \, (1-1/e). \quad (10)$$

We have checked that within the range of parameters studied here, the so-obtained value of $\tau_m$ is the same as the one deduced from the fit of $MSD_m(t)$ by Eq. 2. If we set $f(x) = x^2$ (exp(1/x) + exp(-1/x) – 2), Eqs. 10 and 8 give

$$\frac{\tau_m}{T} = x\left(1 + \ln\frac{f(x)}{g(x)}\right). \quad (11)$$

When x is large, or $\tau \gg$ T, the previous equation reads $\tau_m = \tau + T/3$ at the first order in 1/x. We have checked that this approximation remains excellent as long as x ≥ 1/3. For any x < 1/3, as for instance in the case of a very large diffusion coefficient D (see for example (1)), we have checked by numerical simulation that $\tau_m$ < 2T/3. Thus any *measured* $\tau_m$ ≥ 2T/3 ensures that x ≥ 1/3 and that the real corrected $\tau$ ≥ T/3 is given by

$$\tau = \tau_m - T/3 \quad (12)$$

which is of immediate practical interest to extract equilibration times from measured ones. The real L is then calculated using Eq. 9 which reads

$$L = L_m\left(2\frac{\tau}{T} - 2\left(\frac{\tau}{T}\right)^2\left(1-e^{-\frac{T}{\tau}}\right)\right)^{-1/2} \quad (13)$$

and the real D can now be recovered from Eq. 6. The corrected parameters are thus simply determined using Eqs. 13 and 6. In the case of an exposure (or averaging) time T smaller than the time T' between two successive frames, the calculation leading to Eq. 8 is unchanged as well as its consequences. In particular, Eqs. 12 and 13 still hold. If the confinement geometry is more complex than a square, the whole analysis, based on the correlator in Eq. 5 itself independent of geometry, remains valid. The only difference is that L does then not measure the side of a square but a typical domain size.

To summarize, we have quantified how time-averaging affects observables of biological interest. However, if $\tau$ is large compared to the exposure time T, then $\tau$, L and D remain essentially unmodified. This point is illustrated in both Table 1 and in the following example.

The cases of domain-to-domain jumps or other mechanisms leading to slow long-term diffusion (with coefficient $D_{MAC}$) superimposed to confined short-term diffusion (4) deserve attention (1). We consider, as an example, the movement of the μ-opioid receptor at the surface of a Normal Rat Kidney fibroblast cell from Ref. (3). We calculate $MSD_m(t)$ from one trajectory acquired at common video rate *i.e.* T = 40 ms (see Fig. 1).

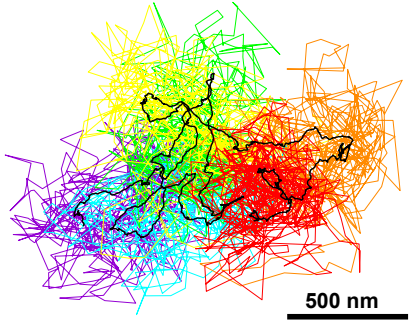

**Fig 1: Experimental trajectory (120 s) of a μ-opioid receptor at the surface of a Normal Rat Kidney cell (3), truncated in successive 20 s segments shown in different colors (indigo, cyan, green, yellow, orange, red) in order to highlight the displacement of the confining domain of size L ≅ 500 nm, in which the receptor diffuses rapidly. The black trajectory shows the slow diffusion of the barycenter of this domain calculated on sliding 4 s intervals.**

Fitting this $MSD_m(t)$ with the generic form:

$$MSD_m^{fit}(t) = \frac{L_m^2}{3}(1 - e^{-\frac{t}{\tau_m}}) + 4D_{MAC}t \qquad (14)$$

we find $\tau_m$ = 192 ms and $L_m$ = 483 nm. We deduce $\tau$ = 178 ms (from Eq. 12), L = 501 nm (from Eq. 13) and finally D = 0.117 μm²s⁻¹ whereas $D_m$ = 0.101 μm²s⁻¹ (from Eq. 6). The influence of time-averaging is weak in this case, as well as in all the trajectories of (3), because T<<$\tau_m$. This validates *a posteriori* the measures in (3). By contrast if $\tau_m$ is smaller than a few T, the corrections to $\tau$ and L must imperatively be taken into account. In addition to the existing numerical data from Ref (1), we have performed numerical simulations in all relevant ranges of the parameters ($\tau \geq$ T/3). The agreement with our analytical predictions is excellent (see Table 1). Our formulae allow to recover accurately the *real* L and D from those measured within a few percents.

**TABLE 1** Comparison of apparent analytically and numerically calculated L and D, and corrected ones, to their real values for different temporal regimes

| $\tau$/T | $L_m$/L | $L_{m,s}$/L | $L_c$/L | $D_m$/D | $D_{m,s}$/D | $D_c$/D |
|---|---|---|---|---|---|---|
| 10 | 0.984 | 0.984 | 1.000 | 0.936 | 0.933 | 0.996 |
| 6.4 | 0.975 | 0.958* | ND | 0.903 | ND | ND |
| 2 | 0.923 | 0.924 | 1.000 | 0.730 | 0.721 | 0.983 |
| 1 | 0.858 | 0.859 | 0.998 | 0.552 | 0.542 | 0.967 |
| 0.5 | 0.753 | 0.755 | 1.002 | 0.341 | 0.342 | 1.002 |
| 0.333 | 0.675 | 0.676 | 1.026 | 0.228 | 0.235 | 1.136 |
| 0.0048 | 0.10 | 0.13* | ND | 0.010 | ND | ND |

$\tau$ is the equilibration time and T the detector exposure time. Parameters without index are real ones. The index m denotes an analytically calculated apparent parameter, m,s a numerically calculated one. The index c denotes a corrected value (ideally equal to the real one) obtained from the numerically calculated one using Eqs. 12, 13 and 6.
* data from Figs. 2 and 3C of Ref (1).
ND: Not Determined because $D_{2-4}$ measured in Ref (1) cannot be used in the present framework.

To conclude, we have demonstrated that the drawbacks of single molecule tracking techniques due to time-averaging are limited. In the case of confined diffusion in membrane domains, we have proposed simple formulae to recover the real domain sizes and diffusion coefficients from those measured. The accuracy remains excellent for confinements with characteristic diffusion times down to $\tau$ = T/3 where T is the exposure time, *i.e.* $\tau$ is of the order of 10 ms at common video rates. Interestingly, this work has shown that events occurring at a time scale smaller than the exposure time can be explored by single molecule tracking.